\documentclass[aps, prx, amsmath,amssymb, reprint, superscriptaddress]{revtex4-2}

\usepackage{graphicx}
\usepackage{bm}
\usepackage{tabularx}
\usepackage{graphicx}
\usepackage{siunitx}
\usepackage{booktabs}
\usepackage{xcolor}
\usepackage[colorlinks=true,citecolor=blue]{hyperref}
\usepackage{natbib}
\usepackage{tablefootnote}

\begin{document}
	
	\title{Semimetal Contacts to Monolayer Semiconductor: \\Weak Metalization as an Effective Mechanism to Schottky Barrier Lowering}

	
	\author{Tong Su}
    \affiliation{Sauvage Laboratory for Smart Materials, School of Materials Science and Engineering, Harbin Institute of Technology, Shenzhen 518055, China}
    \affiliation{Shenzhen Key Laboratory of Flexible Printed Electronics Technology, Harbin Institute of Technology, Shenzhen 518055, China}	\affiliation{Science, Mathematics and Technology, Singapore University of Technology and Design, Singapore 487372}
	
	\author{Yueyan Li}
    \affiliation{School of Information Management, Nanjing University, Nanjing 210023,China}	
    \affiliation{Jiangsu Key Laboratory of Data Engineering and Knowledge Service, Nanjing 210023, China}

    \author{Qianqian Wang}
    \affiliation{Science, Mathematics and Technology, Singapore University of Technology and Design, Singapore 487372}	
    
    \author{Weiwei Zhao}
    \affiliation{Sauvage Laboratory for Smart Materials, School of Materials Science and Engineering, Harbin Institute of Technology, Shenzhen 518055, China}
    \affiliation{Shenzhen Key Laboratory of Flexible Printed Electronics Technology, Harbin Institute of Technology, Shenzhen 518055, China}
    
    \author{Liemao Cao}
    \email{Email: liemao\_cao@hynu.edu.cn}
    \affiliation{College of Physics and Electronic Engineering, Hengyang Normal University, Hengyang 421002, China}

	\author{Yee Sin Ang}
	\email{Email: yeesin\_ang@sutd.edu.sg}
	\affiliation{Science, Mathematics and Technology, Singapore University of Technology and Design, Singapore 487372}	

\begin{abstract}

Recent experiment has uncovered semimetal bismuth (Bi) as an excellent electrical contact to monolayer MoS$_2$ with ultralow contact resistance. 
The contact physics of the broader semimetal/monolayer-semiconductor family beyond Bi/MoS$_2$, however, remains largely unexplored thus far. 
Here we perform a comprehensive first-principle density functional theory investigation on the electrical contact properties between six archetypal two-dimensional (2D) transition metal dichalcogenide (TMDC) semiconductors, i.e. MoS$_2$, WS$_2$, MoSe$_2$, WSe$_2$, MoTe$_2$ and WTe$_2$, and two representative types of semimetals, Bi and antimony (Sb). 
As Bi and Sb work functions energetically aligns well with the TMDC conduction band edge, Ohmic or nearly-Ohmic $n$-type contacts are prevalent.
The interlayer distance of semimetal/TMDC contacts are significantly larger than that of the metal/TMDC counterparts, which results in only weak metalization of TMDC upon contact formation.
Intriguingly, such weak metalization generates semimetal-induced gap states (MIGS) that extends below the conduction band minimum, thus offering an effective mechanism to reduce or eliminate the $n$-type Schottky barrier height (SBH) while still preserving the electronic structures of 2D TMDC. 
A modified Schottky-Mott rule that takes into account SMIGS, interface dipole potential, and Fermi level shifting is proposed, which provides an improved agreement with the DFT-simulated SBH. 
We further show that the tunneling-specific resistivity of Sb/TMDC contacts are generally lower than the Bi counterparts, thus indicating a better charge injection efficiency can be achieved through Sb contacts.
Our findings reveal the promising potential of Bi and Sb as excellent companion electrode materials for advancing 2D semiconductor device technology. 

\end{abstract}

\maketitle

\section{Introduction}

Two-dimensional (2D) semiconductors offer an exciting material platform for extending the legacy of Moore's law beyond silicon \cite{https://doi.org/10.1002/adma.202109894, huang20222d, akinwande2019graphene}. 
Their ultimately-thin body enables excellent electrostatic control of the channel, which, when combined with the exceptional electrical properties, provides a route towards ultrascaled sub-10-nm field-effect transistor (FET) technology that are promising in delivering a future generation of computing electronics with simultaneous high performance and ultralow energy consumption. 
The path towards high-performance 2D semiconductor devices is, however, severely impeded by multiple technical roadblocks \cite{https://doi.org/10.1002/adma.202109796, thomas2021industry}.
In particular, the design and fabrication of high-efficiency Ohmic contact to 2D semiconductor represents one of the major challenges \cite{ZHENG2021100298, https://doi.org/10.1002/inf2.12168}.
The search of a CMOS compatible Ohmic contact engineering approach remains an ongoing challenge thus far.
%
\begin{table*}[htbp]
\centering
\label{tab:3}  
\caption{\textbf{Summary of DFT calculated physical quantities of the 12 types of semimetal/TMDC contacts.} $E_b$ is the binding energy, $\epsilon$ is the lattice mismatch, $d$ is the perpendicular interlayer distance, $\varepsilon_F^\text{(iso)}$ is the Fermi level of the isolated semimetal after receiving a strain arising from lattice mismatch, $\varepsilon_F^\text{(sm/sc)}$ is the Fermi level of the contact heterostructure, $E_\text{EA}$ is the electron affinity, $E_\text{IP}$ is the ionization potential, $\Phi_\text{SBH}^{(e)}$ is the electron ($n$-type) Schottky barrier height, $\Delta V$ is the interface potential difference, $\Delta_\text{M}$ is the weak metalization energy window below conduction band edge, $\Delta \varepsilon_F$ is the Fermi level shifting of the semimetal upon forming contact heterostructures, $\Phi_\text{t}$ is the van der Waals gap tunneling potential height and $w_\text{t}$ is the width of the tunneling potential width. The unstrained isolated Bi and Sb slabs have a Fermi level of -4.12 eV and -4.25 eV, respectively, below the vacuum layer. The stacking configurations are quoted in the format of `semimetal super cell, TMDC supercell'. All energy-related quantities ($E_b$, $\varepsilon_F^\text{(iso)}$, $\varepsilon_F^\text{(sm/sc)}$, $E_\text{EA}$, $E_\text{IP}$, $\Phi_\text{SBH}^{(e)}$, $\Delta V$, $\Delta_\text{M}$, $\Delta \varepsilon_F$ and $\Phi_\text{t}$) are in the unit of (eV) and all length-related quantities ($d$ and $w_\text{t}$) are in the unit of {\AA}.}
\resizebox{\textwidth}{!}{
\begingroup
\begin{tabular}{lccccccccccccccc}
\hline\hline\noalign{\smallskip}
Heterostructures	&	$E_b$	&	$\epsilon$ ($\%$)	&	Stacking 	&	$d$	&	$\varepsilon_F^\text{(iso)}$	&	$\varepsilon_F^\text{(sm/sc)}$	&	$E_\text{EA}$	&	$E_\text{IP}$	&	$\Phi_\text{SBH}^{(e)}$	&	$\Delta V$	&	$\Delta_\text{M}$	&	$\Delta \varepsilon_F$	&	$\Phi_\text{t}$	&	$w_\text{t}$ \\
\noalign{\smallskip}\hline\noalign{\smallskip}
MoS$_2$/Bi	&	-2.41	&	3.8	&	$2\times$2, $3\times3$	&	2.94	&	-4.33	&	-4.45	&	-4.33	&	-5.99	&	0.08	&	0.17	&	0.05	&	0.21	&	2.80	&	1.26\\
WS$_2$/Bi	&	-2.22	&	3.8	&	$2\times2$, $3\times3$ &	3.03	&	-4.33	&	-4.29	&	-3.95	&	-5.75	&	0.13	&	0.03	&	0.22	&	0.21	&	2.97	&	1.33\\
MoSe$_2$/Bi	&	-1.49	&	4.6	&	$2\times2$, $\sqrt{7}\times\sqrt{7}$	&	3.08	&	-3.86	&	-4.03	&	-3.95	&	-5.42	&	-0.09	&	-0.15	&	0.16	&	-0.26	&	2.68	&	1.30\\
WSe$_2$/Bi	&	-1.42	&	4.6	&	$2\times2$, $\sqrt{7}\times\sqrt{7}$ &	3.15	&	-3.86	&	-3.88	&	-3.62	&	-5.23	&	-0.09	&	-0.02	&	0.35	&	-0.26	&	2.75	&	1.42\\
MoTe$_2$/Bi	&	-2.36	&	2.2	&	$2\times2$, $\sqrt{7}\times\sqrt{7}$	&	3.14	&	-4.31	&	-4.24	&	-3.87	&	-4.90	&	0.00	&	0.21	&	0.37	&	0.19	&	2.58	&	1.26\\
WTe$_2$/Bi	&	-2.24	&	2.0	&	$2\times2$, $\sqrt{7}\times\sqrt{7}$	&	3.21	&	-4.30	&	-4.23	&	-3.66	&	-4.71	&	0.04	&	0.33	&	0.54	&	0.18	&	2.65	&	1.26\\
MoS$_2$/Sb	&	-1.03	&	3.8	&	$2\times2$, $\sqrt{7}\times\sqrt{7}$	&	3.08	&	-4.00	&	-4.34	&	-4.34	&	-6.02	&	-0.08	&	-0.29	&	0.08	&	-0.25	&	3.09	&	1.47\\
WS$_2$/Sb	&	-0.98	&	3.8	&	$2\times2$, $\sqrt{7}\times\sqrt{7}$	&	3.21	&	-4.00	&	-4.12	&	-3.94	&	-5.76	&	-0.06	&	-0.08	&	0.24	&	-0.24	&	3.15	&	1.50\\
MoSe$_2$/Sb	&	-1.51	&	3.8	&	$2\times2$, $\sqrt{7}\times\sqrt{7}$	&	3.07	&	-4.38	&	-4.32	&	-3.95	&	-5.37	&	-0.06	&	0.10	&	0.43	&	0.13	&	2.96	&	1.33\\
WSe$_2$/Sb	&	-1.45	&	3.8	&	$2\times2$, $\sqrt{7}\times\sqrt{7}$	&	3.16	&	-4.38	&	-4.31	&	-3.62	&	-5.04	&	0.06	&	0.21	&	0.63	&	0.13	&	3.14	&	1.41\\
MoTe$_2$/Sb	&	-4.00	&	2.5	&	$3\times3$, $\sqrt{13}\times\sqrt{13}$	&	3.26	&	-4.10	&	-4.13	&	-3.87	&	-4.86	&	-0.04	&	0.15	&	0.30	&	-0.15	&	2.73	&	1.27\\
WTe$_2$/Sb	&	-3.91	&	2.7	&	$3\times3$, $\sqrt{13}\times\sqrt{13}$	&	3.32	&	-4.09	&	-4.11	&	-3.66	&	-4.69	&	-0.03	&	0.24	&	0.48	&	-0.16	&	2.82	&	1.35\\
\noalign{\smallskip}
\hline\hline
\end{tabular}
\endgroup}
\end{table*}

\begin{figure*}
    \centering
    \includegraphics[scale=0.858]{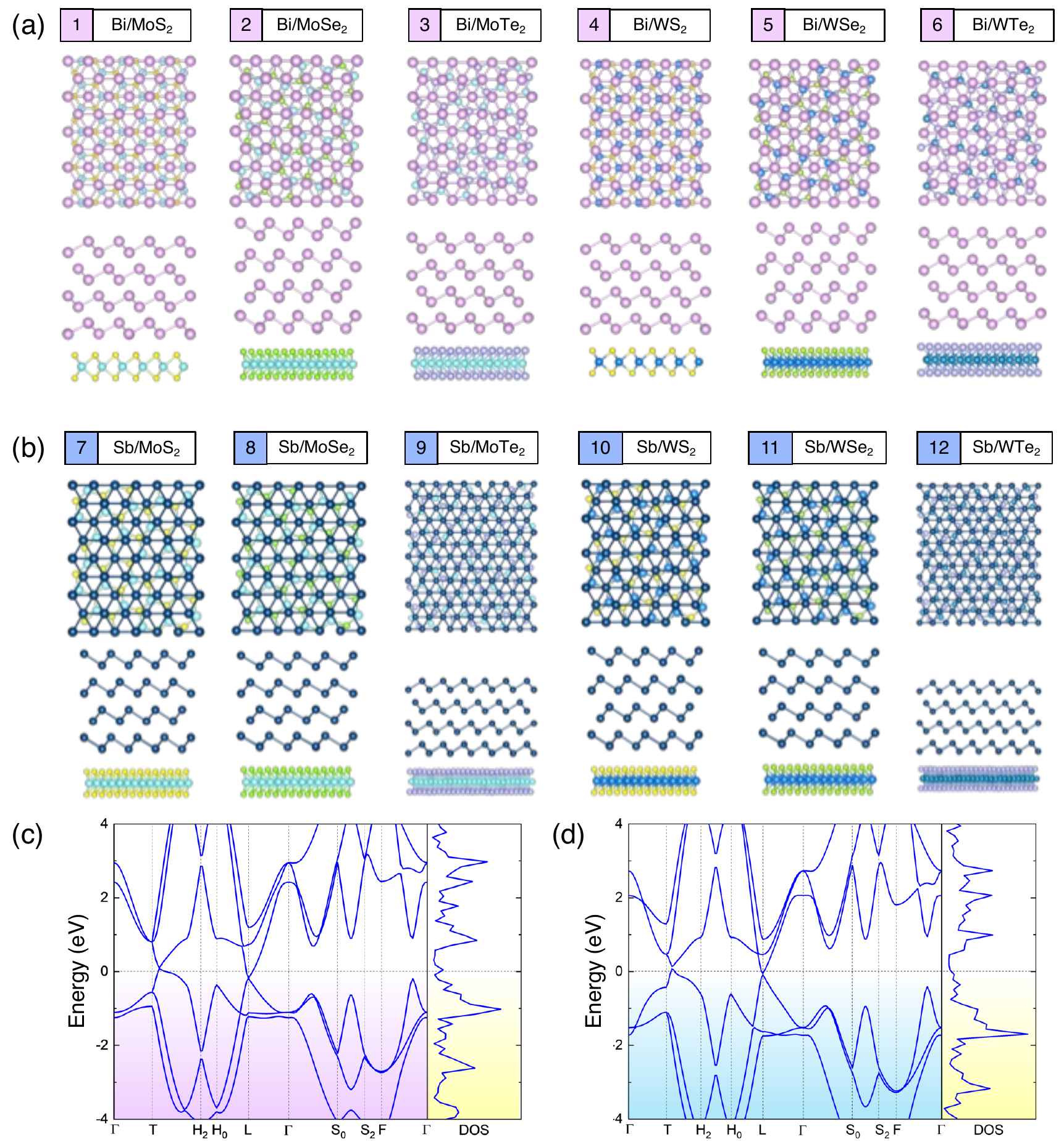}
    \caption{\textbf{Stacking configurations of semimetal contacts to monolayer transition metal dichalcogenide (TMDC) and band structures of semimetals.} (a) Stacking configurations of bismuth-contacted monolayer TMDC. (b) Same as (a) but with antimony (Sb) contact. (c) and (d) show the band structures and electronic density of states of isolated Bi and Sb bulk.  }
    \label{fig:1}
\end{figure*}

\begin{figure*}
    \centering
    \includegraphics[scale=0.5258]{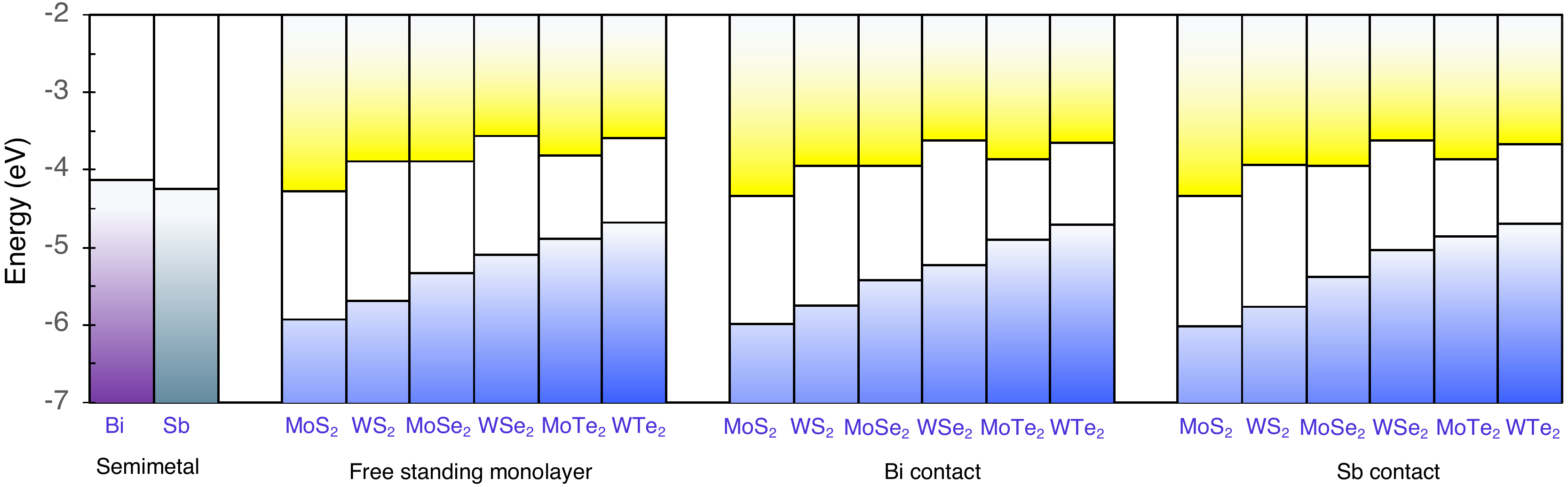}
    \caption{\textbf{Band alignment of semimetals, monolayer TMDCs and the semimetal-contacted TMDCs.} The Fermi level of semimetals and the band edge energies of the monolayer TMDCs are measured with respect to vacuum level. }
    \label{fig:2}
\end{figure*}

Recently, semimetal electrodes, such as bismuth (Bi), antimony (Sb) and tin (Sn), emerge as an excellent contact candidate materials to tranisiton metal dichalcogenide (TMDC) monolayer semiconductor with an experimentally demonstrated low Schottky barrier height (SBH), low contact resistance ($R_\text{c}$) and high on-state current ($I_\text{on}$) \cite{shen2021ultralow, chou2021antimony, chou2020high, doi:10.1063/1.5094890}. 
$R_\text{c}$ as low as 123 $\Omega\mu$m has been achieved in Bi/MoS$_2$ contact, with a sizable on-state current density of $I_\text{on} \sim 10^3 \mu$A$/\mu$m. 
Semimetals thus represent a promising CMOS-compatible strategy in resolving the contact engineering challenges of 2D semiconductor devices, which is in contrast to the van der Waals (vdW) 2D and 3D contact engineering approaches that typically involve stringent fabrication processes not readily compatible with the mainstream CMOS technology \cite{doi:10.1021/acs.chemrev.1c00735, liu2019van, zheng2021ohmic}.
Instead of relying on weak interfacial coupling to tame the Fermi level pinning (FLP) effect -- an undesirable effect that pins the Fermi level ($\varepsilon_F$) of an electrode to the midgap region of a semiconductor \cite{PhysRev.71.717} which leads to an undesirably high contact barrier that cannot be easily reduced \cite{kim2017fermi, gong2014unusual, liu2022fermi} -- as in the case of 2D and 3D vdW contacts \cite{liu2016van, liu2018approaching, wang2019van, huang2019tunable, yu2021electronic, JIANG2019122}, the limited availability of states around the Fermi level ($\varepsilon_F$) in semimetallic Bi intrinsically quenches FLP upon contact formation. 
The weakened FLP enables the $\varepsilon_F$ of Bi to be more freely align to the conduction band minima (CBM) of MoS$_2$, thus achieving Ohmic or nearly Ohmic contacts with ultralow contact resistance.
Recent demonstrations of dual-gated Bi/WS$_2$ FET device geometry \cite{9834930,9706475} show that $R_\text{c}$ can be improved by nearly 20 times via contact gating while multiscale modelling predicts an exceedingly low $R_c<100$ $\Omega\mu$m when the doping level of TMDC and an appropriate dielectric environment are chosen \cite{doi:10.1063/5.0097213}. 
These results suggest the availability of multiple tuning knobs that can be used to further optimize the contact resistance, thus unravelling the enormous potential of semimetal-based 2D semiconductor device technology \cite{quhe2021sub}.

Although spearheading experimental efforts focusing on several semimetal/monolayer-TMDC contacts have shed important light on the device level performance of semimetal contacts, the interfacial electronic properties and contact formation at the microscopic physics level, which are important for device design and computational modelling , remains largely incomplete thus far.
Whether semimetals can be employed to the broader family of 2D TMDC for achieving efficient contacts remains an open research question that needs to be urgently answered before the full potential of semimetal contact engineering can be fully harnessed. 

In this work, we carry out first-principle density functional theory (DFT) investigations that comprehensively covers 12 species of semimetal/semiconductor contact heterostructures composed of transition metal dichalcogenide (TMDC) monolayers $MX_2$ ($M = $ Mo, W; $X = $ S, Se, Te) \cite{jariwala2014emerging} and semimetals Bi and Sb. 
We show that $n$-type Ohmic and nearly Ohmic contacts with ultralow Schottky barrier height (SBH) prevail among the 12 contact heterostructures due to the presence of weak metalization `tail' states that extend below the CBM \cite{PhysRevX.4.031005}. 
The weak metalization induced by Bi and Sb thus provides an effective mechanism to achieve efficient Ohmic contact engineering for TMDC monolayers without introducing severe FLP effect.
A modified Schottky-Mott relation with inclusion of (i) interface dipole correction; (ii) Fermi level offset of semimetal upon contact formation; and (iii) a hybridization term characterizing the degree of hybridization between semimetals and monolayer semiconductors is proposed, which provides an improved agreement with the observed SBH when compared to the original Schottky-Mott relation.
We further show that Sb is a potentially superior semimetal contact due to their generally lower tunneling-specific resistivity than that of the Bi counterparts. 
Our findings provide practical insights on the electronic structures and contact properties of semimetal/2D-TMDC contact heterostructures, and shall pave a foundation for the computational design and the experimental exploration of high-performance low-power 2D semiconductor devices based on semimetal electrical contacts.

\section{Computational Methods}

First-principle calculations are performed with the Vienna ab initio simulation package (VASP) with projector augmented wave method \cite{kresse1993ab,kresse1996efficiency,blochl1994projector,kresse1999ultrasoft}.The exchange-correlation interaction with the generalized gradient approximation (GGA) of Perdew, Burke and Ernzerhof (PBE) \cite{perdew1996generalized} is set and the DFT-D3 \cite{grimme2010consistent} vdW better describe the interlayer vvdW interaction. A kinetic energy cutoff of 500 eV to make the plane expansion meet the $10^{-2}$ eV/ {\AA} force and $10^{-6}$ eV energy convergence criteria. A fine Monkhorst-Pack k-sampling is set in the gamma-centered Brillouin zone with $6 \times 6 \times 1$. We apply a vacuum spacing of at least 15 {\AA} to the heterostructure. Dipole corrections are applied due to the symmetry-breaking nature of the contact heterostructures \cite{neugebauer1992adsorbate}. 

\section{Results and Discussion}

\subsection{Structural and Lattice structures}

The stacking configurations of the Bi and Sb contacts to TMDC monolayers are shown in Fig. 1(a) and 1(b), respectively. 
We employ the combinations of $2\times 2$ Bi with $3 \times 3$ (Mo,W)S$_2$, and $3\times 3$ Sb with $\sqrt{13} \times \sqrt{13}$ (Mo,W)Te$_2$, while the rest of the contact heterostructures are all based on $2\times 2$ semimetal combined $\sqrt{7} \times \sqrt{7}$ TMDC monolayer (see Table I for summary of the stacking configurations and other calculated quantities).
Four layers of Bi or Sb atoms are used to simulate the bulk semimetal contacts. 
The lattice mismatch lies within the range of 2\% to 4.6\%, and the strains are applied to the semimetals so to preserve the electronic band structures of monolayer TMDCs.  
The fully relaxed interlayer distance of the contacts are typically $>3$ {\AA}, which is significantly larger than those of the electrical contacts between 3D metals and TMDC monolayer \cite{wang2021efficient, doi:10.1063/5.0117670, https://doi.org/10.1002/adts.201900001}. 
Such sizable interlayer distance spatially decouple the semimetals and the monolyaer TMDC, thus leads to a diminished metalization between semimetal and monolayer TMDC.  
The binding energies of the heterostructures are calculated as
\begin{equation}
E_\text{b} = E_\text{sm/sc} - E_\text{sm} - E_\text{sc}
\end{equation}
where $E_\text{sm/sc}$, $E_\text{sm}$ and $E_\text{sc}$ are the total energy per supercell of the contact heterostructure, the semimetal and the semiconductor, respectively.
The negative-valued $E_\text{b}$ of all heterostructures indicate that the contact formation is energetically favourable (see Table I). 
The band structure of bulk Bi and Sb are shown in Figs. 1(c) and 1(d). 
The conduction and valance bands lightly touch or overlap near the Fermi level at $L$ point and along the $T$-$H_2$ axis, thus yielding the semimetallic nature. 

\subsection{Electronic properties: Band structure and interfacial electron redistribution}

\begin{figure*}
    \centering
    \includegraphics[scale=0.858]{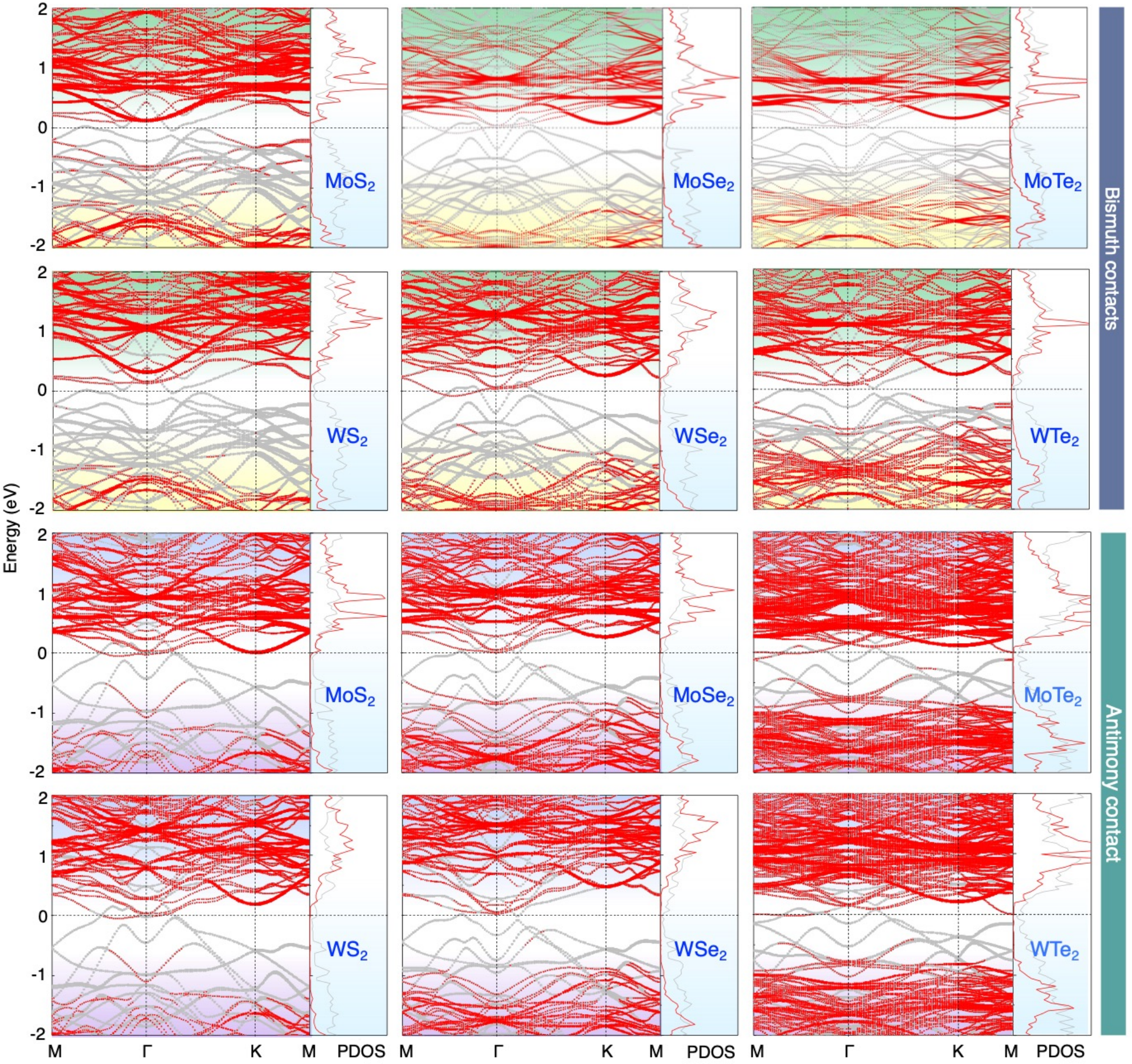}
    \caption{\textbf{Band structures of semimetal-contacted monolayer TMDC.} The band structures are projected onto semimetal (grey) and monolayer semiconductor (red). The projected density of states (PDOS) are shown in the left panel of each band structure plots. }
    \label{fig:3}
\end{figure*}

The band alignment of TMDC monolayers and the $\varepsilon_F$ of Sb and Bi are shown in Fig. 2. 
The $\varepsilon_F$ of Bi and Sb (i.e. 4.12 eV and 4.25 eV below the vacumm level, respectively) aligns at the close proximity of the CBM of monolayer TMDC ($\sim 4$ eV).
Such band alignment thus suggests monolayer TMDC to form $n$-type Schottky or Ohmic contacts with Bi and Sb semimetals. 
However, it should be noted that the Anderson rule cannot be directly applied to determine the contact types due to the presence of interfacial charge transfer and orbital hybridization effect that could significantly modify the band edge energies and the overall band alignment of the contact heterostructures \cite{besse2021beyond}.
This aspect is particularly evident from the electronic band structures of the semimetal contacts shown in Fig. 2. 
Here although the $\varepsilon_F$ of Bi and Sb are energetically separated from the CBM of TMDC other than (Mo,W)S$_2$ monolayers by a sizable $0.2 \sim 0.7$ eV, Ohmic or nearly Ohmic contacts are still prevalent among these semimetal-contacted TMDC monolayers.
As shown below, the presence of weak metalization in semimetal-contacted TMDC is highly beneficial in lowering the SBH and transfers the contacts towards Ohmic or low-SBH contacts. 

\begin{figure*}
    \centering
    \includegraphics[scale=0.858]{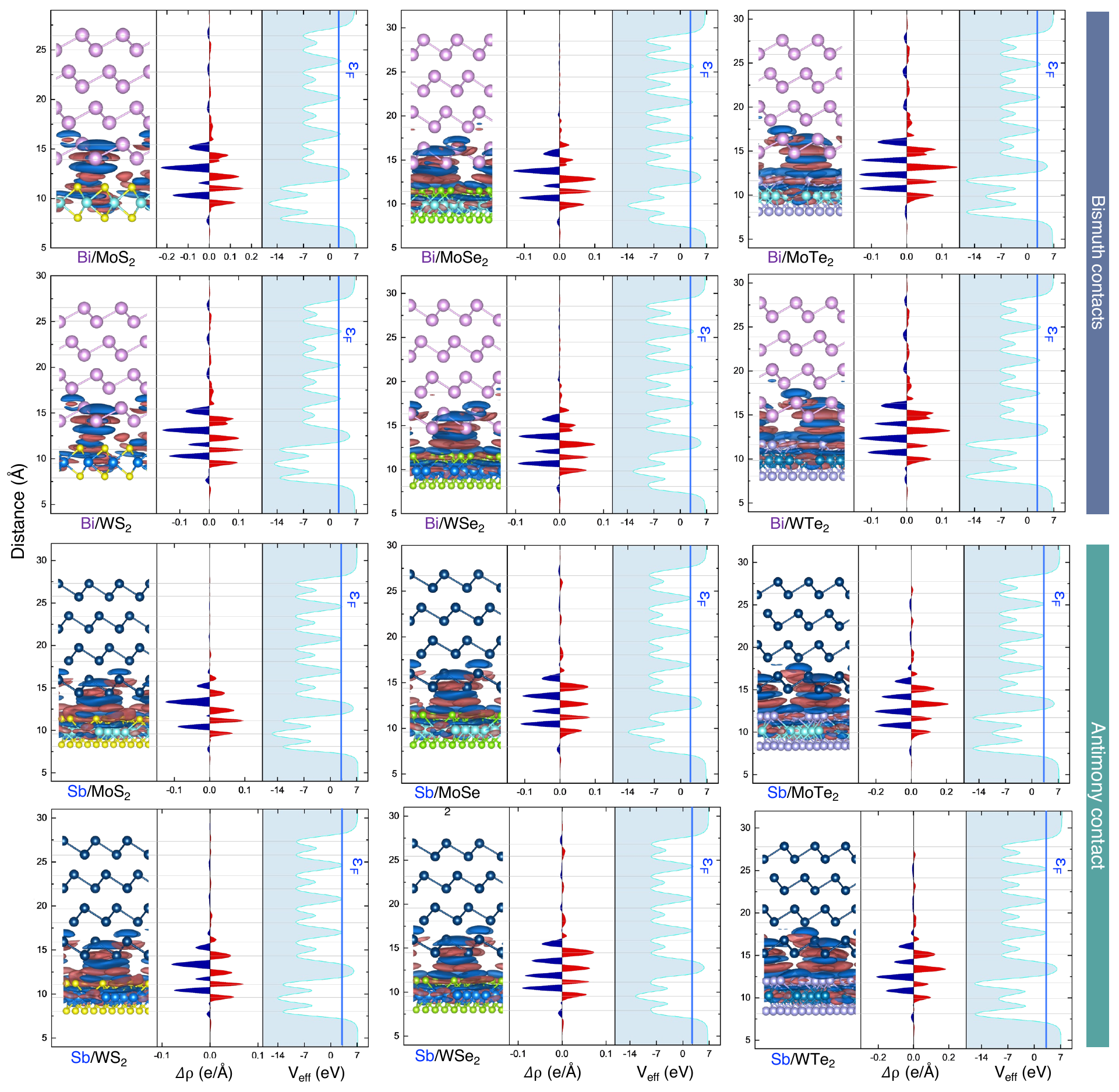}
    \caption{\textbf{Charge transfer and electrostatic potential profile of the semimetal/TMDC contacts.} The differential charge density distribution plotted onto the lattice (left panel), the plane-averaged differential charge density plots (middle panel) and the plane-averaged electrostatic potential profiles (right panel) are shown for Bi-and Sb-contacted TMDC monolayers. }
    \label{fig:4}
\end{figure*}

A closer inspection on the projected band structures and the projected density of states (PDOS) reveals that the Ohmic contact formations in semimetal-contacted TMDC monolayers arises from the weak metalization induced by Bi and Sb upon contact formation.
The presence of weak metalization on TMDC creates semimetal-induced gap states (SMIGS) that extend below the CBM energies of TMDC monolayers, thus closing up the energy mismatch, i.e. SBH, between the semimetal and the monolayer TMDC. 
Weak metalization thus offer a mechanism to lower the SBH while still preserving the semiconducting band edge states of monolayer TMDC. 
Low SBH is particularly beneficial for thermionic and tunneling charge injection \cite{PhysRevB.96.205423, PhysRevLett.121.056802} into monolayer TMDC, thus paving a way to achieve ultralow contact resistance in 2D semiconductor devices.
We further note that the DFT-calculated SBHs of Bi/MoS$_2$ (0.075 eV) and Bi/WS$_2$ (0.13 eV) differ from the experimentally extracted zero-height (i.e. Ohmic) of Bi/MoS$_2$ and 0.040 eV of Bi/WS$_2$. 
We suspect that such discrepancy could arise from the inevitable presence of defects at the contacted region in experiment which leads to stronger metalization and more extensive SMIGS in monolayer TMDC that lower the SBH of the contacts. 
Nevertheless, the general trend of Bi/WS$_2$ having a larger SBH than Bi/MoS$_2$ obtained via DFT calculations is agreement with that observed experimentally. 

The electron redistribution and the interfacial potential difference reveal great wealth of contact physics and interfacial interactions of metal/semiconductor heterostrcutures \cite{PhysRevB.90.201411, PhysRevB.93.085304, doi:10.1063/5.0010849}.
The electron redistribution across the contact interface can be assessed via the differential charge density of the heterostructures ($\Delta \rho$), which is can be calculated as
\begin{equation}
    \Delta \rho = \rho_{\text{sm}/\text{sc}}  - \rho_{\text{sm}} - \rho_{\text{sc}}
\end{equation}
where $\rho_{\text{sm}/\text{sc}}$, $\rho_{\text{sm}}$ and $\rho_{\text{sc}}$ are the electron densities of the contact heterostructure, semimetal and semiconductor, respectively.
Electron redistribution across the interface results in the formation of an interface potential difference, i.e.
\begin{equation}
    \Delta V = W_{\text{sm}/\text{sc}}^\text{(sm)} - W_{\text{sc}/\text{sc}}^\text{(sc)}
\end{equation}
where $W_{\text{sm}/\text{sc}}^{(sc)}$ and $W_{\text{sc}/\text{sc}}^{(sc)}$ are the work functions extracted from the semiconductor and the semimetal sides of the contact heterostructure, respectively. 

The $\Delta \rho$ and $\Delta V$ are shown in Fig. 4. 
We observe complex electron redistribution patterns across the contact interfaces from the $\Delta \rho$ plots (see left and middle panels of Fig. 4), in which electron accumulation (depletion) are denoted by red (blue) isosurfaces and red (blue) shaded curves. 
Charge redistribution occurs through the entire thickness of TMDC due to the weak screening effect of 2D semiconductor, which is in stark contrast to that of semimetals in which electron redistribution effect are rapidly screened and does not penetrate beyond the outermost layer of Bi or Sb atoms. 
In general, electrons accumulate on both (i) the outermost contacting layer of the semimetals; and (ii) the chalcogen atoms at the contact interface.
These electron accumulations arise from different physical origins. 
The electron accumulation on the semimetal atoms is due to the electronic pushback (or `pillow') effect \cite{PhysRevB.90.201411, PhysRevB.93.085304}. 
In this case, electrons are pushed back towards the semimetal so to minimize the wavefunction overlap, and hence the Pauli exchange repulsion, between the semimetal and the chalcogen atoms at the contact interface. 
This results in electron accumulation on the semimetal sides of the contact interface.
The electron accumulation on the chalcogen atoms occurs due to their stronger electronegativity nature when compared with the semimetal atoms.
The stronger electronegativity of the chalcogen atoms causes a net displacement of electron densities towards TMDC, thus leading to electron accumulation on the chalcogen atoms.
Correspondingly, whether the resulting dipole potential $\Delta V$ is positive-or negative-valued depends on the `tug-of-war' between the electronic pushback effect and the electronegativity difference of semimetal and chalcogen atoms. 
Contact with stronger electronic pushback effect exhibits a positive valued $\Delta V$, and is indicative of physisorptive interfacial interaction \cite{PhysRevB.93.085304}. 
In contrast, if the electron transfer due to electronegativity difference dominates over the electronic pushback effect, a negative valued $\Delta V$ is obtained, which is indicative of chemisorptive interfacial interaction \cite{PhysRevB.93.085304}. 
For semimetal/2D-TMDC contacts, the $\Delta V$ ranges from $-0.29$ to $0.33$ eV, thus suggesting that both chemisorption and physisorption can occur upon contact formation.

\begin{figure}
    \centering
    \includegraphics[scale=1.158]{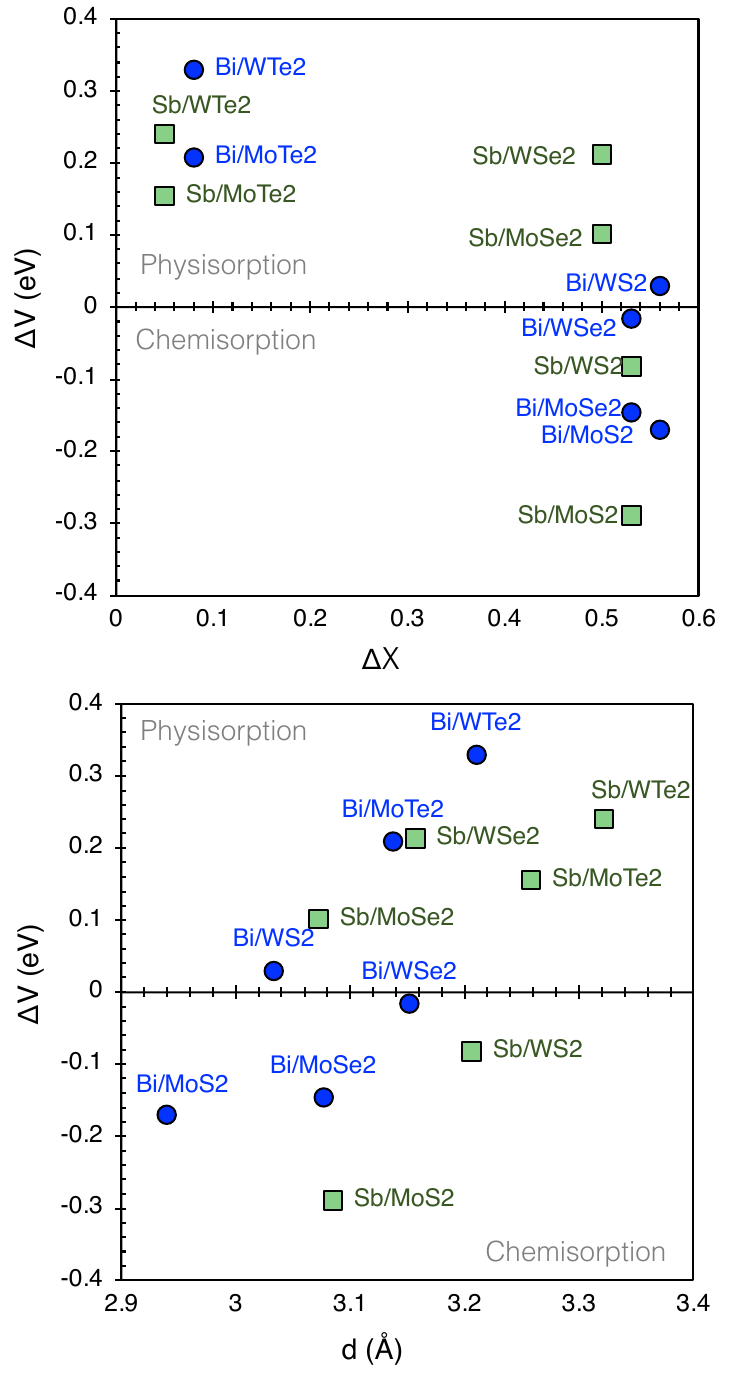}
    \caption{\textbf{Interface dipole potential ($\Delta V$) of semimetal/TMDC contacts.} (a) $\Delta V$ as a function of the electronegativity difference $\Delta \chi$. (b) $\Delta V$ as a function of the interlayer distance $d$.}
    \label{fig:5}
\end{figure}

We define the electronegativity difference ($\Delta \chi$) as
\begin{equation}
    \Delta \chi = \chi_\text{chalcogen} - \chi_\text{semimetal}
\end{equation}
where $\chi_\text{chalcogen}$ and $\chi_\text{semimetal}$ are the electronegativity of the chalcogen and semimetal atoms, respetively.
Here $\chi_\text{S,Se,Te} = (2.58, 2.55, 2.1)$, respectively, and $\chi_\text{Bi,Sb} = (2.02, 2.05)$, respectively. 
Contacts with small $\Delta \chi$ exhibits positive-valued $\Delta V$ [see Fig. 5(a) for a plot of $\Delta V$ versus $\Delta \chi$] due to the dominance of electronic pushback effect which leads to overall charge accumulation on the semimetal. 
The interfacial interaction of semimetal-contacted Te-based TMDC, which has small $\Delta \chi$, is thus dominated by the electronic pushback effect, and the contacts fall under the physisorption type with weaker interfacial coupling effect.
In contrast, contacts with large $\Delta \chi$ tends to exhibit low or negative valued $\Delta V$ [Fig. 5(a)] due to the stronger overall electron density shifting towards the TMDC side. 
Here the interface couples strongly, thus resulting in the chemisorption contact types in S-and Se-based TMDC upon contacted by semimetals. 
The physisorption of Te-based TMDC and chemisorption of S-and Se-based TMDC on semimetals are also evident from the semimetal-TMDC interlayer distance ($d$) [Fig. 5(b)].
Contacts with smaller $d$ tends to have lower or negative-valued $\Delta V$ which is consistent with chemisorptive interfacial interaction that binds the two materials more strongly, while contacts with larger $d$ tends to have larger or positive-valued $\Delta V$ which is indicative of physisorptive interfaical interaction in which electronic pushback effect dominates. 

It should be noted that the physisorption and chemisorption effect are also observed in metal/2D-TMDC contact heterostructures, such as the physisorptive Pt/MoS$_2$ \cite{doi:10.1063/5.0010849} which preserves the electronic structures of TMDC, and the chemisorptive Ti/MoS$_2$ \cite{PhysRevB.93.085304}, In/MoS$_2$ \cite{kim2021origins} and W/MoS$_2$ \cite{gao2020titanium} in which the TMDC is severely metalized. 
In contrast, semimetal/2D-TMDC contact exhibits only mild metalization when compared to that of the metal counterparts \cite{chen2017general} (see band structure and PDOS plots in Fig. 3.) 
This aspect thus represents a major distinction between metal and semimetal contacts to 2D TMDCs. 

\begin{figure}
    \centering
    \includegraphics[scale=1.158]{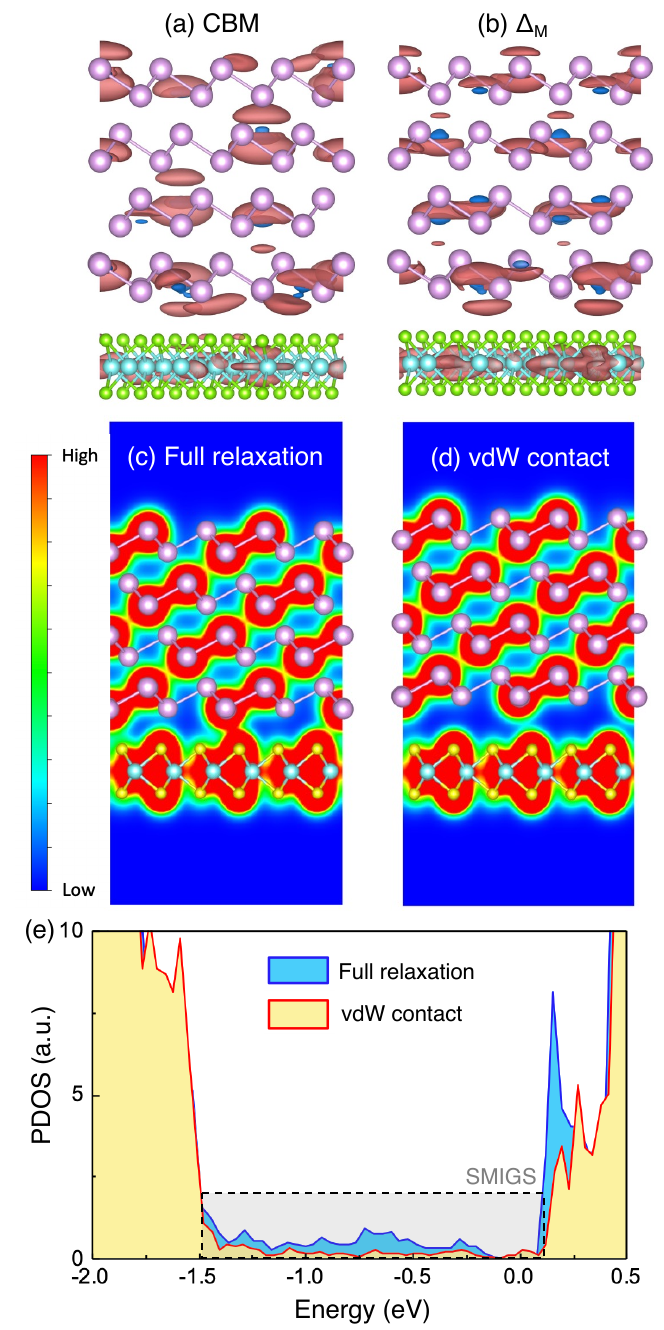}
    \caption{\textbf{Metalization and van der Waals contacts in semimetal/TMDC heterostructures.} (a) The band-decomposed charge density plot of CBM states in Bi/MoSe$_2$ contact. (b) Same as (a) but for the energy range of $\Delta_\text{M}$ below the CBM. (c) Charge density plot of fully relaxed Bi/MoS$_2$ with interlayer distance of 2.94 {\AA}, and (d) with vdW distance 3.86 {\AA}. (e) Electronic density of states projected on to MoS$_2$ monolayer for fully relaxed and vdW contacts. }
    \label{fig:6}
\end{figure}

\subsection{Metalization effect and van der Waals contact}

We now investigate the metalization effect on TMDC upon forming contact with semimetals. 
Taking Bi/MoSe$_2$ as a representative contact, the band-decomposed charge density [see Figs. 6(a) and 6(b)] of the CBM states and within the energy window of $\Delta_\text{M}$ below the CBM which contains the SMIGS as one can directly seen from the band structure and PDOS plots in Fig. 3. 
For CBM states, the charge density spatially distributes in both TMDC and semimetal. 
As $\Delta H$ lies in the band gap regime of TMDC, the charge density should only concentrate on the semimetals but not on the TMDC in the case of an ideal non-interacting semimetal/TMDC contact interface. 
This is, however, not the case in a realistic contact, such as the illustrative example of Bi/MoSe$_2$ shown in Figs. 6(b). 
Here the charge density distribution of the electronic states lying within the $\Delta H$ energy window extends well into MoSe$_2$, thus indicating the presence of metalization effect.

The metalization of TMDC can be suppressed by spatially separating the two contacting materials to reduce the interfacial interactions.
In fact, such strategy has been employed in metal/TMDC contact via the formation of atomically clean and weakly coupled vdW contacts \cite{liu2016van, liu2018approaching, wang2019van, LIU20191426} which offers an effective way to suppress FLP and to achieve ultralow contact resistance. 
We compare the contact properties of a fully-relaxed Bi/MoS$_2$ heterostructure with an interlayer distance $d$ (see Table I) and that of a Bi/MoS$_2$ contact constructed using vdW gap distance ($d_\text{vdW}$) \cite{kwon2022interaction},
\begin{equation}
    d_\text{vdW} = r_\text{vdW}^\text{(Bi)} + r_\text{vdW}^\text{(S)}
\end{equation}
where $r_\text{vdW}^\text{(Bi)} = 2.07$ {\AA} and $r_\text{vdW}^\text{(S)} = 1.87$ {\AA} are the vdW radii of Bi and S atoms, respectively. 
At the fully-relaxed interlayer distance $d$, the electronic orbitals of Bi overlap with that of the S atoms [see Fig. 6(c)]. 
Such orbital overlap is significantly reduced in the case of vdW contact [Fig. 6(d)]. 
The presence of SMIGS in MoS$_2$ upon contacted by Bi can be clearly observed in the PDOS in Fig. 6(e). 
In the case of vdW contact, the SMIGS are significantly suppressed. 
Because of the suppression of the SMIGS in vdW contacts, the PDOS of MoS$_2$ terminates sharply below the CBM. 
This results in a larger SBH in vdW-type Bi/MoS$_2$ contact, thus suggesting that vdW contact may not be ideal for applications where SBH are required to be minimized. 

\begin{figure*}
    \centering
    \includegraphics[scale=1.0858]{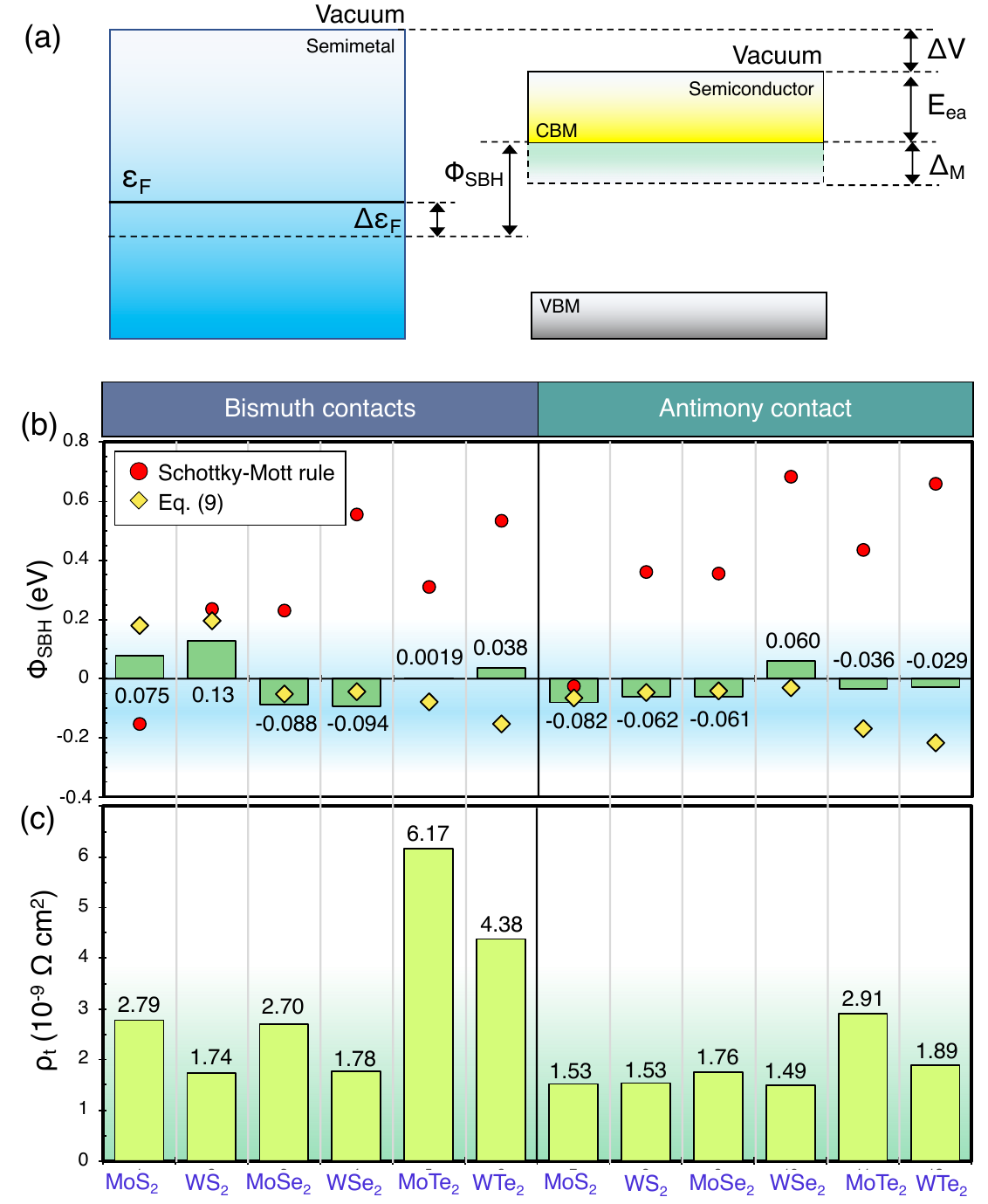}
    \caption{\textbf{Schottky barrier heights (SBH) and tunneling-specific contact resistivity ($\rho_\text{t}$) of semimetal/TMDC contacts.} (a) Schematic drawing of the modified Schottky-Mott (SM) rule that takes into account SMIGS ($\Delta_\text{M}$), interface dipole potential ($\Delta V$) and Fermi level shifting ($\Delta \varepsilon_F$) of the semimetal. (b)SBH extracted from the DFT calculated band structure and PDOS (vertical bars), and the predicted SBH values of SM rule [Eq. (6)] and of modified SM rule [Eq. (9)]. (c) Tunneling-specific contact resistivity calculated based on Simmons tunneling diode model.}
    \label{fig:7}
\end{figure*}

\subsection{Modified Schottky-Mott theory and tunneling-specific resistivity}

For a non-interacting ideal metal/semiconductor contact, the electron SBH ($\Phi_\text{SBH}^\text{(e)}$) of an $n$-type Schottky contact is given by the Schottky-Mott (SM) rule \cite{doi:10.1063/1.4858400},
\begin{equation}
    \Phi_\text{SBH}^\text{(e)} = W_\text{metal} - E_\text{EA}
\end{equation}
where $W_\text{metal}$ and $E_\text{EA}$ are the isolated metal work function and the electron affinity of the isolated semiconductor. 
The ideal SM rule, however, omits the metal-semiconductor interfacial interactions which are inevitable in almost all realistic contact heterostructures \cite{PhysRevB.103.035304, kong2020doping}. 
For 2D-metal/2D-semiconductor contact, the presence of an interfacial potential difference ($\Delta V$) induced by charge redistribution generates an additional band offset term \cite{https://doi.org/10.1002/adts.201700001} that modifies the SBH. Accordingly, the SM rule is modifed as \cite{si2016controllable, ding2021engineering, li2022revealing}
\begin{equation}
    \Phi_\text{SBH}^\text{(e)} = W_\text{metal} - E_\text{EA} - \Delta V
\end{equation}
where a positive $\Delta V$, i.e. electron accumulation on 2D metal, increases the metal work function and lowers the electron SBH. 
In 3D-metal/2D-semiconductor contact, the presence of significant interfacial interaction further complicates the SBH formation physics \cite{allain2015electrical}. 
For example, in Au-contacted MoS$_2$, an intermediate reconstructed Au$_4$S$_4$ layer can form at the Au/MoS$_2$ interface, which modifies the SM rule by an additional quasi-bonding correction term that accounts for the formation of electronic tail states extending from the band edges induced by quasi-bonding (QB) effect \cite{PhysRevB.105.224105}.
The presence of QB-induced gap states (QBIGS) modifies the SM rule to
\begin{equation}
    \Phi_\text{SBH}^\text{(e)} = W_\text{metal} - E_\text{EA} - \Delta V - \Delta_\text{QB}
\end{equation}
where $\Delta_\text{QB}$ is the energy window of the QBIGS. 
Motivated by the formulation of Eq. (8), we propose a modified SM rule for semimetal/2D-TMDC contacts, i.e.
\begin{equation}
    \Phi_\text{SBH}^\text{(e)} = W_\text{metal} - E_\text{EA} - \Delta V + \Delta \varepsilon_F - \Delta_\text{M}
\end{equation}
where $\Delta \varepsilon_F \equiv \varepsilon_{F}^\text{(sm/sc)} - \varepsilon_F^\text{(iso)}$ is the Fermi level shifting upon contact formation which arises due to lattice mismatch when constructing the heterostructure supercell, and $\Delta_\text{M}$ is the energy window of the SMIGS. 
In Fig. 6(a), we compare the electron SBH extracted from PDOS, the predicted SBH from the SM rule in Eq. (6), and that from the modified SM rule in Eq. (9). 
We find that while the original SM rule fails to predict the SBH of almost all contacts, the modifed SM rule exhibits a significantly improved agreement with the DFT calculation results [Fig. 7(a)]. 
This observation thus highlights the importance of both interface potential difference and SMIGS in the formation of SBH in semimetal/2D-TMDC contacts.  

The presence of a vdW gap at the contact interface yields a potential barrier through which the electrons must tunnel across in order to form an injection current into the monolayer TMDC. 
The corresponding \emph{tunneling-specific contact resistivity} \cite{shen2021ultralow, wang2021efficient, doi:10.1063/5.0117670} can be computed based on the Simmons tunneling diode model for a square potential barrier \cite{doi:10.1063/1.1702682}, i.e. 
\begin{equation}
    \rho_\text{t} \approx \frac{4\pi^2\hbar w_\text{t}^2}{e^2} \frac{\exp\left(\frac{2\sqrt{2m}}{\hbar} w_\text{t} \phi_\text{t}^{1/2}\right)}{\frac{\sqrt{2m}}{\hbar} w_\text{t}\phi_\text{t}^{1/2}-1}
\end{equation}
where $\phi_\text{t}$ and $w_\text{t}$ are the vdW gap tunneling barrier height and width measured with respect to the Fermi level, respectively, which can be extracted from the plane-average electrostatic plots across the contact heterostructures (i.e. Fig. 4) \cite{wang2021schottky}. 
The calculated $\rho_t$ [Fig. 7(b)] are generally in the order of $10^{-9}$ $\Omega$cm$^2$, which is similar to that of Bi/MoS$_2$ \cite{shen2021ultralow} and other 2D semiconductor metal contacts, such as MoSi$_2$N$_4$ \cite{wang2021efficient} and $\gamma$-GeSe \cite{doi:10.1063/5.0117670} monolayers. 
The Te-based TMDC generally exhibit larger $\rho_\text{t}$ than other TMDC monolayers due to the physisorptive contact interface with larger interlayer distance, which generates a wider tunneling barrier to suppresses electron tunneling.
We further observe that the Sb/2D-TMDC contacts generally exhibits lower $\rho_\text{t}$ when compared to that of the Bi counterparts. 
The lower $\rho_\text{t}$ and better thermal stability \cite{chou2021antimony} of Sb compared to Bi contacts suggests that Sb contact shall serve as an efficient semimetal contact materials for 2D TMDC.

\section{Conclusion}
In conclusion, we performed a first-principle density functional theory simulation to investigate the contact properties between Bi and Sb semimetal to monolayer TMDCs. 
We found that weak metalization generates semimetal-induced gap states (SMGIS) that are beneficial in lowering the Schottky barrier height (SBH), thus offering a pathway to achieve Ohmic or low-barrier contacts without compromising the semiconductor electronic band structures. 
The physisorptive and chemisorptive nature of the various contacts are identified. 
A modified Schottky-Mott (SM) rule that takes into account interface dipole potential and SMGIS are found to improve the theoretical description of SBH over the original SM rule. 
Finally, the contact-specific resistivity of Sb contacts are found to be generally lower than Bi contacts, thus unravelling the potential of Sb-contacted monolayer TMDC for energy-efficient device application. 
These findings shall offer a theoretical foundation and design blueprint for the future exploration of semimetal-enabled high-performance 2D semiconductor nanoelectronics and novel devices, such as valleytronics \cite{ma1, valley} and multiferroics \cite{ma2, mf}. 

\section*{Data availability statement}
The data that support the findings of this study are available upon reasonable request from the authors.

\section*{Acknowledgement}
This work is funded by the Singapore Ministry of Education (MOE) Academic Research Fund (AcRF) Tier 2 Grant (MOE-T2EP50221-0019) and SUTD-ZJU IDEA Visiting Professor Grant (SUTD-ZJU (VP) 202001). T. S. is supported by the China Scholarship Council (CSC). W. Z. is supported by the Natural Science Foundation of China (No. 52073075) and Shenzhen Science and Technology Program (Grant No. KQTD20170809110344233). The computational work for this article was partially performed on resources of the National Supercomputing Centre, Singapore (https://www.nscc.sg). 


\end{document}